\documentstyle[11pt,newpasp,twoside,psfig]{article}
\index{neutron stars!accretion}
\index{X-ray binaries}\index{binaries: X-ray}

\def\edcomment#1{\iffalse\marginpar{\raggedright\sl#1\/}\else\relax\fi}
\reversemarginpar

\begin{document}
\title{Accretion onto the neutron star in Be/X-ray binaries}
 \author{Kimitake Hayasaki}
\affil{Devision of Physics, graduate school of science, Hokkaido University, Kitaku, Sapporo 060-0810, Japan}
\author{Atsuo. T. Okazaki}
\affil{Faculty of Engineering, Hokkai-Gakuen University, Toyohira-ku, Sapporo 062-8605, Japan}

\begin{abstract}

We study accretion onto the neutron star in Be/X-ray binaries,
using a 3D SPH code and the data imported from a high resolution simulation by Okazaki et al.\ (2002) 
for a coplanar system with a short period ($P_{\rm orb}=24.3\ \rm d$) 
and moderate eccentricity $(\rm{e=0.34})$.
We find that a time-dependent accretion disk 
is formed around the neutron star in Be/X-ray binaries. 
The disk shrinks after the periastron passage of the Be star 
and restores its radius afterwards. 
Our simulations show that the truncated Be disk model
for Be/X-ray binaries is consistent with the observed X-ray behavior.

\end{abstract}

\section{Introduction}

The Be/X-ray binaries represent the largest subclass of high-mass
X-ray binaries. These systems consist of a neutron star 
and a Be star with a cool equatorial disk.
The orbit is wide and usually eccentric.
Most of the Be/X-ray binaries show only transient activity in the X-ray
emission. These outbursts result from the transient accretion onto the neutron star
from the circumsteller matter of the Be star. 
In this paper, we simulate the accretion flow around the neutron star in Be/X-ray binaries,   
using a 3D SPH code (Bate et al. 1995) and the mass-transfer rate from the Be-star disk obtained
by Okazaki et al. (2002).

\section{Non-steady accretion disk around the neutron star}

In Be/X-ray binaries, the mass-transfer rate has strong phase dependence (Okazaki et al. 2002). 
 Therefore, the structure of an accretion
disk formed around the neutron star in these systems is also
likely to be strongly phase dependent. 

Figure 1 gives the radial disk structures and the snapshots
 of a developed accretion disk around the neutron star for $7 \le t/P_{\rm{orb}} < 8$.   
From the figure, we note that the material transferred from the Be-star disk to the neutron star 
forms a non-steady accretion disk around the neutron star. 
It is noted from the upper panels that the disk is nearly
Keplerian at any phase and the accretion flow is highly subsonic
except in the outermost part for a short period of time when
the material of the Be-star disk is transferred to the neutron star.

The disk shrinks at periastron passage
by a negative torque exerted by the Be star. 
In addition, the ram pressure of the supersonic infall of matter
captured by the neutron star around periastron enhances the density
in the outermost part of the accretion disk, making the disk outer
edge sharp. Afterwards, the disk restores its radius by the viscous diffusion.
Thus, our simulation confirms that the strucutre of the accretion disk in Be/X-ray binaries 
has a strong dependence on the orbital phase.

\begin{figure}
\centerline{
\psfig{file=hayasakik1_1.ps,width=3cm}
\psfig{file=hayasakik1_2.ps,width=3cm}
\psfig{file=hayasakik1_3.ps,width=3cm}
\psfig{file=hayasakik1_4.ps,width=3cm}}
\end{figure}
% ===========================================
\begin{figure}
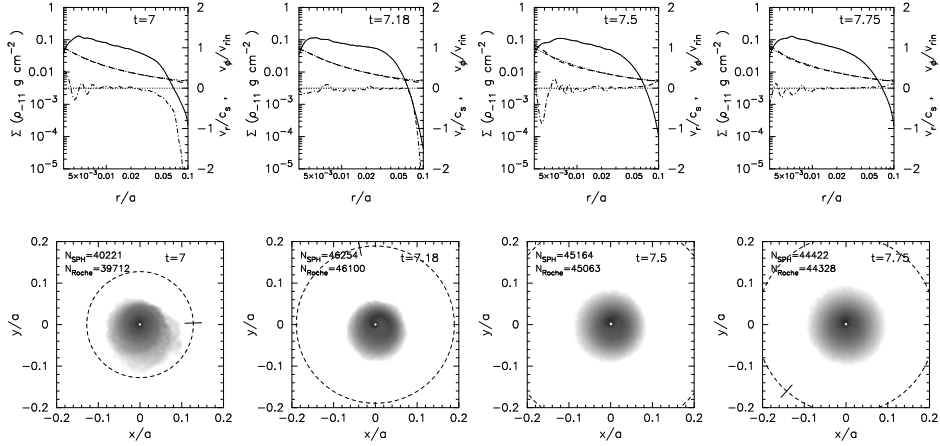

\centerline{
\psfig{file=hayasakik1_5.ps,width=3cm}
\psfig{file=hayasakik1_6.ps,width=3cm}
\psfig{file=hayasakik1_7.ps,width=3cm}
\psfig{file=hayasakik1_8.ps,width=3cm}}
\caption{
Phase-dependent radial structures (upper panels) and snapshots (lower panels) 
of a developed accretion disk around the neutron star for $7 \le t/P_{\rm{orb}} < 8$. 
In the upper panels, the solid, the dot-dashed and the dashed line 
denote the logarithm of the surface density, the radial Mach number and
the azimuthal velocity normalized by the Keplerian velocity 
at $r_{\rm{in}}=3.0\times10^{-3}\rm{a}$, where $\rm{a=6.6\times10^{12}\ cm}$ 
is the semi-major axis, respectively. 
The surface density is measured in units of $\rm{\rho_{-11}\ g\ cm^{-2}}$, 
where $\rm{\rho_{-11}}$ is highest local density 
in the Be-star disk normalized by $\rm{10^{-11}} \rm{g}\ \rm{cm^{-3}}$. 
The lower panels show the logarithm of the surface density. 
The dashed circle denotes the effective Roche lobe of the neutron star.
The short line-segment indicates the direction of the Be star. 
Annotated at the top-left corner of each panel are the number of SPH particles, $N_{\rm SPH}$, 
and the number of particles inside the effective Roche lobe of the neutron star, $N_{\rm Roche}$.
}
\end{figure}

% ===========================================

\acknowledgments
This work was supported in part by Nukazawa Science Foundation.

\end{document}